\documentclass[aps,pre,twocolumn,groupedaddress,showpacs,floatfix]{revtex4}

\usepackage{graphicx}

\begin{document}

%\preprint{}

\title{Directed geometrical worm algorithm applied to the quantum rotor model}


\author{Fabien Alet$^{(1,2)}$}
\email{alet@phys.ethz.ch}
%\thanks{}

\author{Erik~S.~S\o rensen$^{(3)}$}

\affiliation{$^{(1)}$Computational Laboratory, ETH Z\"urich, CH-8092 Z\"urich,
  Switzerland}
\affiliation{$^{(2)}$Theoretische Physik, ETH Z\"urich, CH-8093 Z\"urich, Switzerland}
\affiliation{$^{(3)}$Department of Physics and Astronomy, McMaster University, Hamilton, ON, L8S 4M1 Canada}
\date{\today}

\begin{abstract}
We discuss the implementation of a directed geometrical worm algorithm for the study of quantum link-current models.
In this algorithm
Monte Carlo updates are made through the biased reptation of a worm
through the lattice. A directed algorithm is an algorithm where, during the construction of
the worm, the probability for erasing the immediately preceding part of the worm, when adding a new part,
is minimal. We introduce a simple numerical procedure for minimizing this probability.
The procedure only depends on appropriately defined local probabilities and should be
generally applicable. Furthermore we show how correlation functions, $C({\bf r},{\bf \tau})$ can be straightforwardly
obtained from the probability of a worm to reach a site $({\bf r}, {\bf \tau})$ away from its starting point independent
of whether or not a directed version of the algorithm is used.
Detailed analytical proofs of the
validity of the Monte Carlo algorithms are presented for both the directed and un-directed geometrical
worm algorithms. Results for  auto-correlation times
and Green functions are presented for the quantum rotor model.  

\end{abstract}

\pacs{02.70.Ss, O2.70.Tt, 74.20.Mn}
%\keywords{}

\maketitle


%%%%%%%%%%%%%%%%%%%%%%%%%%%%%%%%%%%%%%%%%%%%%%%%%%%%%%%%%%%%%%%%%%%%%%%%%%
%%%%%%%%%%%%%%%%%%%%%% Introduction %%%%%%%%%%%%%%%%%%%%%%%%%%%%%%%%%%%%%%
%%%%%%%%%%%%%%%%%%%%%%%%%%%%%%%%%%%%%%%%%%%%%%%%%%%%%%%%%%%%%%%%%%%%%%%%%%

\section{Introduction}

Improving and developing new numerical algorithms lies at the heart of
computational physics.  Amongst others, Monte Carlo (MC) methods are often seen
as the best choice for the study of phase transitions taking place in classical
or quantum models. For the study of spin models for example, cluster
algorithms, either in the classical~\cite{Swendsen,Wolff} or
quantum~\cite{Evertz,Sandvik,Prokofev} case, perform non-local moves in phase
space, allowing for the treatment of systems much larger than with traditional local
update methods (single spin-flip algorithms). These types of algorithms have almost completely
solved the problem of critical slowing down arising near phase transitions.

The class of systems for which cluster methods are known to exist is
limited, and it is therefore of great interest to search for new algorithms
possessing the same efficient features for other models. In this context, we
have proposed recently a non-local ''worm'' algorithm for the study of
quantum link-current models~\cite{Alet03}. These models arise from a phase
approximation of bosonic Hubbard models, but are also relevant in the context
of quantum electrodynamics~\cite{Banks}. Previous MC simulations of the
quantum link-current (quantum rotor) model used a local algorithm suffering
from critical slowing down. In the new
algorithm~\cite{Alet03} updates are made by reptating a ''worm'' through
the lattice~\cite{Prokofev,Prokofev.Classical}. Since the movement of the worm
only depends on a few probabilities determined locally with respect to the
current position of the ``head" of the worm, we call this type of algorithm
a geometrical worm algorithm as opposed to other recently developed worm algorithms
based on high-temperature series expansions~\cite{Prokofev.Classical}.
The geometrical worm algorithm gives rise to very small
autocorrelation times and by directing the algorithm these autocorrelation times
can be even further reduced.

In this paper, we briefly recall the principles of the geometrical worm
algorithm~\cite{Alet03}.  During the construction of a worm a new part is added to the worm
by moving the worm through one of the $\sigma$ nearest neighbor links. Usually the
associated $\sigma$ probabilities, $p_\sigma$, are chosen in an un-biased geometrical way and
there is therefore a significant probability that the new part of the worm
will back-track in its own path, thereby erasing the immediately preceding
part. In many cases this back-tracking (or bounce) probability is the dominant
probability among the $\sigma$ probabilities and using these un-biased probabilities
is therefore clearly rather wasteful.  Here we describe an improvement of this
geometric worm algorithm, which we call the directed worm algorithm, as a reference
to recently developed directed loop methods for Quantum Monte Carlo simulations of spin systems~\cite{Sylju}.
This directed geometrical worm algorithm is identical to
its un-directed counterpart except for the fact that the probabilities $p_\sigma$
are now chosen in a biased way, using knowledge of the immediately preceding step in the construction
of the worm. These biased probabilities can all be tabulated at the start of the
simulation and the additional computational effort stems solely from the significantly
wider distribution of the directed worms.
The directed algorithm gives rise to even better results, as
will be shown in the following part of this paper, where we present results
on autocorrelation times for both directed and ``undirected'' worm algorithms.
The procedure for choosing the ``biased" probabilities leading to the directed
algorithm is quite general and should be applicable to other algorithms that depend
on local probabilities.  Furthermore, we show how Green functions $C({\bf r},\tau)$ of the
original quantum model can be measured efficiently during the construction of
the worm by calculating the probability that the worm reaches a given site $({\bf r}, \tau)$
away from its starting point, independently of whether a directed or un-directed algorithm is
used. For both the derivation of the directed algorithm and the
measurements of correlation functions, analytical proofs of the validity of the
algorithms are presented.

The outline of the paper is as follows: in the next section, we present the
quantum rotor model and introduce some useful notation. Then, a brief
description of the ''undirected'' worm algorithm is given in
section~\ref{sec:undirected}, before proceeding to the  main contents of this
paper, a description of the directed geometrical worm algorithm
(section~\ref{sec:directed}). A simple procedure for numerically determining
the biased probabilities for the worm moves is presented. In addition, we
derive a proof of detailed balance for the directed worm algorithm.  In order
to compare our algorithms to related ones, we present in
section~\ref{sec:classical} another recent approach due to Prokof'ev and
Svistunov~\cite{Prokofev.Classical}, originally based on high temperature
series expansion for classical statistical models, which we therefore will
refer to as "classical worms" throughout this paper.  In section~\ref{sec:auto},
we estimate the efficiency of the algorithms by calculating auto-correlation
times in the MC simulation, and compare to both undirected and classical
algorithms. We then discuss the measurements of correlation functions within
the worm algorithm in section~\ref{sec:cfunc} and show some results at a
specific point of the phase diagram. We conclude with a discussion of the
features of the directed algorithm.

%%%%%%%%%%%%%%%%%%%%%%%%%%%%%%%%%%%%%%%%%%%%%%%%%%%%%%%%%%%%%%%%%%%%%%%%%%
%%%%%%%%%%%%%%%%%%%%%%%%%%%%% Model %%%%%%%%%%%%%%%%%%%%%%%%%%%%%%%%%%%%%%
%%%%%%%%%%%%%%%%%%%%%%%%%%%%%%%%%%%%%%%%%%%%%%%%%%%%%%%%%%%%%%%%%%%%%%%%%%

\section{The model}
Many magnetic systems, Josephson Junction arrays and several other systems
can be described by a quantum rotor model~\cite{Sachdev}:
\begin{equation}
H_{\text{qr}}=
\frac{U}{2}\sum_{\bf r}
\left( \frac{1}{i}\frac{\partial}{\partial
\theta_{\bf r}} \right)^2
+i\sum_{\bf r} \mu_{\bf r}
\frac{\partial}{\partial \theta_{\bf r}}
-t\sum_{\langle {\bf r},{\bf r'}\rangle }
\cos(\theta_{\bf r}-\theta_{\bf r'}).
\label{eq:hqr}
\end{equation}
Here,
$\theta_{\bf r}$ is the phase of the quantum rotor, $t$ the renormalized
coupling strength and $\mu_{\bf r}$ an effective chemical potential. If
$\mu\equiv 0$ it can be shown that this model displays the same critical
behavior as the $D+1$ dimensional $XY$-model. However, when $\mu_{\bf r}\neq 0$
this model is not amenable to direct numerical treatment in this representation
due to the resulting imaginary term.
It is therefore very noteworthy that an equivalent {\it completely real} representation in terms
of link-currents exists even for non-zero $\mu_{\bf r}$.
This link-current (Villain) representation is a classical (2+1)D equivalent Hamiltonian that is usually written in
the following manner~\cite{Sorensen}: 
\begin{equation}
H=\frac{1}{K} {\sum_{({\bf r},\tau)}}
\left[\frac{1}{2} {\bf J}_{({\bf r},\tau)}^{2}- \mu_{\bf r} J_{({\bf r},\tau)}^\tau\right].
\label{eq:hV}
\end{equation}
The sum is taken over all divergenceless current configurations ${\bf
\nabla \cdot J} = 0$. The degrees of freedom are ``currents'' ${\bf
J}=(J^x,J^y,J^\tau)$ living on the links of the
lattice. These link variables $J^x,J^y,J^\tau=0,\pm 1, \pm 2, \pm 3 \ldots$
are integers. $K$ is the
effective temperature, varying like $t/U$ in the quantum rotor model. We refer to Ref.~\cite{Sorensen} for a 
precise derivation of this model and for a description of its physical implications.

Another incentive for studying the critical behavior of the quantum rotor model
comes from the close relation between this model and bosonic systems.
Bosonic systems with strong correlations are often described in terms
of the (disordered) boson Hubbard model:
$
H_{\rm bH}=\sum_{\bf r}\left(\frac{U}{2}\hat n_{\bf r}^2 -
\mu_{\bf r}\hat n_{\bf r}\right)- t_0\sum_{\langle {\bf r},{\bf r'}\rangle }
(\hat \Phi^\dagger_{\bf r}\hat \Phi_{\bf r'}+c.c) \ .
$
The correlations are here described by $U$ the on-site repulsion. The hopping strength is given
by $t_0$  and
$\mu_{\bf r}$ the chemical potential varying
uniformly in space between $\mu\pm\Delta$.
$\hat n_{\bf r}=\hat \Phi^\dagger_{\bf r}\hat \Phi_{\bf r}$ is the number operator.
If we set $\hat \Phi_{\bf r}\equiv|\hat \Phi_{\bf r}|e^{i\hat \theta_{\bf r}}$ and
integrate out amplitude fluctuations, it can be shown that $H_{\rm bH}$ is equivalent
to the quantum rotor model~\cite{Sorensen}. For systems where amplitude fluctuations
can be neglected at the critical point, such as granular superconductors and Josephson junctions
arrays, the quantum rotor model should therefore correctly describe the underlying quantum critical phenomena.

In the following we only discuss the quantum rotor model in $d=2$ dimensions corresponding
to the $d+1$ dimensional link-current model. In general a non-zero $\mu$ will introduce separate
dynamics for the time and space directions from which a dynamical critical exponent, $z$ can
be defined. If the divergence of the spatial correlation length close to
criticality is characterized by the exponent $\nu$, $z$ is defined by requiring that
the correlation length in the time direction
diverges with the exponent $z\nu$. For what we will be discussing here $\mu=0$ and $z=1$.

\section{Algorithms}

%%%%%%%%%%%%%%%%%%%%%%%%%%%%%%%%%%%%%%%%%%%%%%%%%%%%%%%%%%%%%%%%%%%%%%%%%%
%%%%%%%%%%%%%%%%%%%%%%%%%%%%% Undirected Algo %%%%%%%%%%%%%%%%%%%%%%%%%%%%
%%%%%%%%%%%%%%%%%%%%%%%%%%%%%%%%%%%%%%%%%%%%%%%%%%%%%%%%%%%%%%%%%%%%%%%%%%

\subsection{The Geometrical (Undirected) Worm algorithm}
\label{sec:undirected}

The quantum rotor model has been extensively studied in the link-current
representation using conventional Monte Carlo technique using local
updates~\cite{Sorensen,Cha,vanOtterlo,Kisker,Park}.

\begin{figure}
\includegraphics[width=8cm]{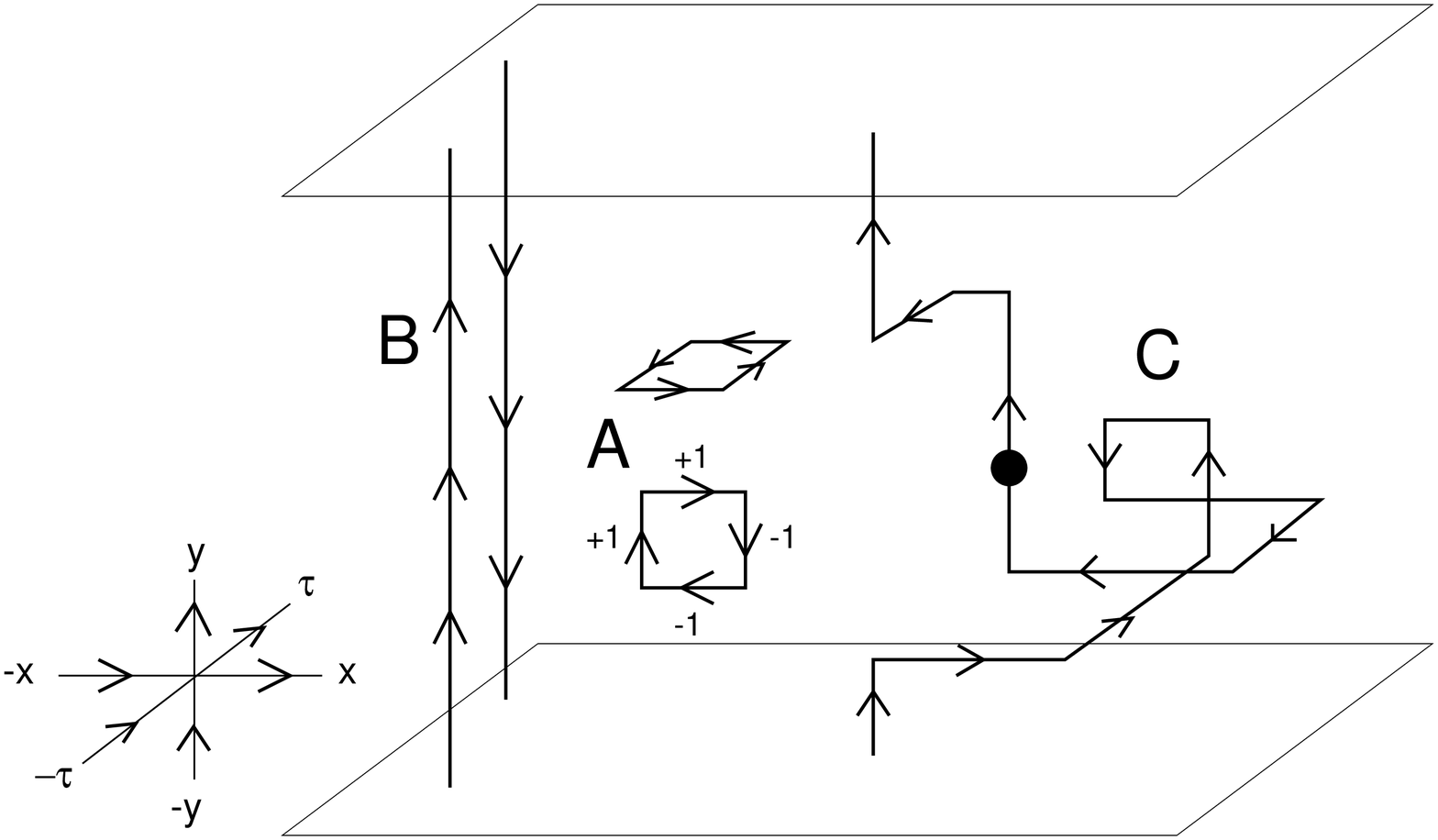}
\caption{Monte Carlo moves in the model. To ensure the divergenceless
condition, only closed moves can be performed. On the left part of the
figures, previous Monte Carlo updates with the local algorithm are
depicted. On the right part is given an example of a move with the worm
algorithm starting from an initial random site (black dot).}
\label{fig:moves}
\end{figure}

Conventional Monte Carlo updates on the model~(\ref{eq:hV}) consists of
updating simultaneously four link variables as shown in the left part of
Fig.~\ref{fig:moves} (A). To ensure ergodicity, one also has to use global
moves, updating a whole line of link variables (B in
Fig.~\ref{fig:moves}). The acceptance ratio for these global moves becomes
exponentially small with the system size for large systems.
Many interesting quantities such as the stiffness, necessary
for the determination of the critical point, and current-current correlations,
necessary for the calculation of transport properties such as the resistivity
and the compressibility, are {\it only} non-zero when these global moves are successful.
An effective Monte Carlo sampling of these global moves is therefore imperative
and it is easy to understand that the performance of the local algorithm is
rather poor, especially near a phase transition: critical slowing down in
the Monte Carlo simulations prohibits the study of large system sizes.

In order to be able to correctly describe the directed geometrical worm
algorithm we have to review the geometrical worm-cluster algorithm introduced in
reference~\cite{Alet03} in some detail. This algorithm allows for non-local moves as the ones
depicted on the right part of Fig.~\ref{fig:moves} (C). The performances of
this algorithm have been reported in the previous work~\cite{Alet03}. This algorithm is closely
related to other cluster algorithms~\cite{Swendsen,Wolff,Evertz} and especially to
''worm'' algorithms~\cite{Prokofev,Sandvik,Prokofev.Classical}, from which we have borrowed the
name. We stress that this algorithm is different from the ''classical'' worm algorithm presented
in Ref.~\cite{Prokofev.Classical} in the sense that it is geometrical: link variables are
not ``flipped" with a thermodynamic probability, instead, a new part of the worm is added by
selecting a direction according to $\sigma$ locally determined probabilities $p_{\sigma}$.
Since these probabilities only depend on the local environment we call them ``{\it geometrical}"
probabilities. Secondly, even though the local probabilities $p_{\sigma}$ do depend on the
effective temperature, $K$, we always have $\sum_\sigma p_\sigma=1$ (per definition) and the only effect
of the temperature is therefore to preferentially move the worm in one direction as opposed to
another one. In some sense this is very similar to the N-fold way~\cite{Bortz} of performing
Monte Carlo simulations.

We now describe the contents of the geometrical algorithm: we update the configurations
by moving a ``worm'' through the lattice of links. The links through which
the worm pass are updated {\it during} its construction. The configurations
generated during the construction (``reptation") of the worm are not valid (the divergenceless
of ${\bf J}$ is not fulfilled) but at the end of its path, when the worm forms a closed loop, this condition is
verified and the final configuration is valid.

We first define a convention for the orientation of the
lattice. Around each site with coordinates $({\bf r}=(x,y),\tau)$, there are
six links on which the integer currents $J^\sigma_{({\bf r},\tau)}$  are defined with
$\sigma=x,y,\tau,-x,-y,-\tau$. The change in $J^\sigma_{({\bf r},\tau)}$ that the
worm will perform during its course depends on whether $\sigma$ is an
incoming or outgoing link: here our convention is to consider positive
$x,y,\tau$ as outgoing directions and $-x,-y,-\tau$ as incoming (see left lower
part of Fig.~\ref{fig:moves}).

If the worm is leaving the site $({\bf r},\tau)$ passing through an outgoing link $\sigma=x,y,\tau$, then 
\begin{equation}
J^\sigma_{({\bf r},\tau)} \rightarrow J^\sigma_{({\bf r},\tau)}+1.
\end{equation}
If it is leaving through an incoming link $\sigma=-x,-y,-\tau$ , we have
\begin{equation}
J^\sigma_{({\bf r},\tau)} \rightarrow J^\sigma_{({\bf r},\tau)}-1.
\end{equation}

Graphically, the convention means that the update in $J^\sigma_{({\bf
r},\tau)}$ is $+1$ ($-1$) if the worm goes in the same (opposite) direction
as the arrows denoted in the left lower part of figure \ref{fig:moves}.

The construction ("reptation") of the worm can be described the following way: first, we start
the worm at a random site $N_1=({\bf r}_1,\tau_1)$ of the lattice (black dot
in figure \ref{fig:moves}).
From this site, the worm has the possibility to
go to one of its six neighboring sites. To choose which direction to take,
a weight $A^\sigma$ is calculated for the six directions $\sigma=\pm x, \pm
y, \pm \tau$. For  $A^\sigma$, we use a Metropolis-like weight:
\begin{equation}
A^\sigma=\min(1,\exp(-(E'_\sigma-E_\sigma)/K))
\label{eq:A}
\end{equation}
where $E_\sigma=\frac{1}{2}(J^\sigma)^2-\mu J^\sigma \delta_{\sigma,\rm
\tau}$ is the local energy carried by the link $\sigma$ and
$E'_\sigma=\frac{1}{2}(J^\sigma \pm 1)^2-\mu (J^\sigma \pm 1)
\delta_{\sigma,\rm \tau}$ is the local energy on the link $\sigma$ {\it if
the worm passes through this link}. The plus or minus sign depends on the
incoming or outcoming nature of the link (see above). Please note that there
are other possible choices for $A_\sigma$~\cite{AletPhD}.

Once the $A_\sigma$'s are calculated, one computes the probabilities
$p_\sigma$ by normalizing the
weights $A_\sigma$ :
\begin{equation}
p_\sigma=\frac{A_\sigma}{N}
\label{eq:Prob}
\end{equation}
where $N=\sum_\sigma A_\sigma$ is the normalization. A random number
uniformly distributed in $[0,1]$ is generated and a direction $\sigma$ chosen
according to equation~(\ref{eq:Prob}). Once a direction is chosen, the
corresponding link variable $J^\sigma$ is updated by $\pm 1$ and the worm
moved to the next lattice site in this direction.

From there, we apply the same procedure to choose another site, modify the
link variable, move the worm until the worm eventually reaches its
starting point and forms a closed loop. This is then the end of this non-local move.

To satisfy the detailed balance condition, this worm move must either be accepted
or rejected. To check this, one has to store the initial and final normalizations
$N_{s_1}^i$ and $N_{s_1}^f$ (calculated as in Eq.~\ref{eq:Prob}) of the weights at
the site $s_1=({\bf r}_1,\tau_1)$. $N_{s_1}^i$ is the initial
normalization {\it before} the worm is inserted and $N_{
s_1}^f$ the final normalization {\it after} the worm reaches the initial
point. The worm move is then accepted with probability $N_{
s_1}^i/N_{s_1}^f$. If the move is rejected, we have
to cancel all changes of the link-currents made during the construction
of the worm. During a typical simulation the rejection probability is
usually very small.

As already mentioned, the link configurations generated during the worm
move do not satisfy the divergenceless constraint, but it is easy to see that
the final configuration does. It is important to note that the worm may pass
many times though the same link and that at each step, it can bounce back (back-track) to
the previous lattice site in its path.

A proof of detailed balance for this algorithm is obtained by considering
the moves of the worm and of an anti-worm, going exactly in the opposite
direction~\cite{Alet03}. This worm algorithm satisfies ergodicity since the worm
can make local loops and line moves as in the local algorithm, which is ergodic.

All in all, the geometrical un-directed worm algorithm can be summarized using
the following pseudo-algorithm:
\begin{enumerate}
\item  Choose a random initial site
$s_1=({\bf r}_1,\tau_1)$ in the space-time lattice.
\item For each of the directions
$\sigma=\pm x,\pm y,\pm \tau$, calculate the weights
$A^\sigma_{s_i}$ with
$A^\sigma_{s_i}=$
$\min(1,\exp(-\Delta E_{s_i}^{\sigma}/K)),$
$\Delta E_{s_i}^{\sigma}=E'^\sigma_{s_i}-E_{s_i}^{\sigma}.$\label{Uloop}
\item Calculate the normalization $N_{s_i}=\sum_\sigma A^\sigma_{s_i}$
and the associated probabilities
$p^\sigma_{s_i}={A^\sigma_{s_i}}/{N_{s_i}}.$
\item According to the probabilities, $p^\sigma_{s_i}$,
choose a direction $\sigma$.
\item Update the $J^\sigma_{s_i}$ for the direction chosen
and move the worm to the new lattice site $s_{i+1}$.
\item If $s_i\ne s_1$ goto \ref{Uloop}.
\item Calculate the normalizations $\bar N_{s_1}$ and $N_{s_1}$
of the initial site, $s_1$, with and without the worm present. Erase
the worm with probability $P^e=1-\min(1,N_{s_1}/{\bar N_{s_1}})$.

\end{enumerate}

%%%%%%%%%%%%%%%%%%%%%%%%%%%%%%%%%%%%%%%%%%%%%%%%%%%%%%%%%%%%%%%%%%%%%%%%%%
%%%%%%%%%%%%%%%%%%%%%%%%%%%%%%% Directed Algo %%%%%%%%%%%%%%%%%%%%%%%%%%%%
%%%%%%%%%%%%%%%%%%%%%%%%%%%%%%%%%%%%%%%%%%%%%%%%%%%%%%%%%%%%%%%%%%%%%%%%%%

\subsection{The Directed Geometrical Worm algorithm}
\label{sec:directed}

The above algorithm for geometrical worms is not optimal since the worm
quite often will choose to erase itself by returning to the previous site.
While it is in general not possible to always set this back-tracking (or bounce)
probability to zero it is quite straightforward to choose the probabilities $p^\sigma_{s_i}$
such that the bounce or back-tracking probability will be eliminated in almost all cases
and in general will be as small as possible. The procedure for doing this amounts to solving
a simple linear programming optimizing problem. If we consider models with disorder this has
to be done at each site, but the correctly optimized (biased) probabilities $p^\sigma_{s_i}$
can still be tabulated at the start of the calculation.

In order to see how we can minimize the back-tracking probability let us define the $6\times 6$ matrix $P_{s_i}$ of 
probabilities where the element $P_{s_i}^{kl}$ of the matrix $P_{s_i}$ is given by the conditional probability
$p_{s_i}(\sigma_k | \sigma_l)$ for going in the direction $\sigma_k$ at the site $s_i$
if the worm is coming from the direction $\sigma_l$. The back-tracking
probabilities at the site $s_i$ now
correspond to the {\it diagonal} elements of the matrix $P_{s_i}$. For the algorithm described in the
previous section $p_{s_i}(\sigma_k | \sigma_l)$ was simply chosen as $A^{\sigma_k}_{s_i}/{N_{s_i}}$
independent of $\sigma_l$. Thus all the columns of $P_{s_i}$ were the same and $P_{s_i}$ had in
general rather large diagonal elements. However, as we shall see below, the matrix $P_{s_i}$ only
needs to satisfy the following two conditions in order to define a working
geometrical worm algorithm. These conditions are:
\begin{eqnarray}
\sum_k p_{s_i}(\sigma_k | \sigma_l)&=&1\ \ {(\rm probability)}\label{eq:prob}\\
\frac{P_{s_i}^{kl}}{P_{s_i}^{lk}}\equiv\frac{p_{s_i}(\sigma_k | \sigma_l)}{p_{s_i}(\sigma_l | \sigma_k)}&=&
\frac{A^{\sigma_k}_{s_i}}{A^{\sigma_l}_{s_i}}\ \ {(\rm detailed\ balance)}\label{eq:balance}.
\end{eqnarray}
These conditions are not very restrictive and will in most cases allow us to define a matrix $P_{s_i}$ with
all the diagonal elements (back-tracking probabilities) equal to zero.
The conventional geometrical worm algorithm, discussed in the previous
section, corresponds to
$P_{s_i}^{kl}=A^{\sigma_k}_{s_i}/{N_{s_i}}$.

If we define a function $f$ as the sum of the diagonal elements of
$P_{s_i}$, $f=\sum_k P_{s_i}^{kk}$, we can reformulate the search for a
matrix $P_{s_i}$ with minimal diagonal elements as a standard linear
programming problem. Writing $P_{s_i}^{kk}=1-\sum_{l\neq k}P_{s_i}^{kl}$
we should minimize $f$ subject to the constraints
$\sum_{l\neq k}P_{s_i}^{kl}\leq 1\ \ \forall k$. The minimum can be found
using standard techniques of linear programming~\cite{numrec} and corresponds in almost all
cases to $f=0$. 
The matrix $P_{s_i}$ depends on the value of all the 6 link-currents
$J^\sigma_{s_i}$. During the construction of the worm only
sites where
$J_{s_i}^{-x}+J_{s_i}^{-y}+J_{s_i}^{-\tau}-J_{s_i}^x-J_{s_i}^y-J_{s_i}^\tau=1$,
$s_i\neq s_1$ occur. Since in general $|J^\sigma_{s_i}|$ will almost never
exceed a certain value $J_{\rm max}$ it is easy to construct a lookup table for the matrices
$P_{s_i}$ at the beginning of the simulation and only calculate
$P_{s_i}(\{J^\sigma_{s_i}\})$ during the simulation if for some $\sigma$
$|J^\sigma_{s_i}|>J_{\rm max}$.

This idea of minimizing the bounce processes is also at the heart of Quantum
Monte Carlo directed loop methods~\cite{Sylju}. The previous restrictions on
the matrix $P$ and the way to solve them numerically are indeed very general,
and constitute a simple framework for how one can construct a directed
algorithm out of a ''standard'' non-local loop, worm or cluster algorithm. 

We can now define a {\it directed} geometrical worm
algorithm with minimal back-tracking probability.
Using a pseudo-code notation we have:
\begin{enumerate}
\item  Choose a random initial site
$s_1=({\bf r}_1,\tau_1)$ in the space-time lattice.
\item {\bf If} $i=1$ {\bf then:}
For each of the directions
$\sigma=\pm x,\pm y,\pm \tau$, calculate the weights 
$A^\sigma_{s_i}$ 
with $A^\sigma_{s_i}=\min(1,\exp(-\Delta E_{s_i}^{\sigma}/K)),$ $
\Delta E_{s_i}^{\sigma}=E'^\sigma_{s_i}-E_{s_i}^{\sigma}$.
Calculate the normalization $N_{s_i}=\sum_\sigma A^\sigma_{s_i}$
and the associated probabilities $ p^\sigma_{s_i}={A^\sigma_{s_i}}/{N_{s_i}}.$
{\bf Else:}
According to the incoming direction, $\sigma_l$, set
$p^\sigma_{s_i}$ equal to the $l$'th column of $P_{s_i}$.
\label{loop}
\item According to the probabilities, $p^\sigma_{s_i}$, 
choose a direction $\sigma$.
\item Update $J^\sigma_{s_i}$ for the direction chosen
and move the worm to the new lattice site $s_{i+1}$.
\item If $s_i\ne s_1$ goto \ref{loop}.
\item Calculate the normalizations $\bar N_{s_1}$, of the site $s_1$ with
the worm present, and $N_{s_1}$, without the worm. 
Erase
the worm with probability $P^e=1-\min(1,N_{s_1}$/$\bar N_{s_1}$).
\end{enumerate}

Now we turn to the proof of detailed balance for the directed
algorithm.  Let us consider the case where the
worm, $w$, visits the sites $\{s_1\ldots s_N\}$ where $s_1$ is the initial
site. The worm then goes through the corresponding link variables
$\{l_1\ldots l_N\}$, with $l_i$ connecting $s_{i}$ and $s_{i+1}$.
Note that $s_N$ is the last site visited before the worm reaches
$s_1$. Hence, $s_N$ and $s_1$ are connected by the link $l_N$.
The total probability for constructing the worm $w$ is then given by:
\begin{equation}
P_w = P_{s_1}(1-P_{w}^e)
\frac{A_{s_1}^{\sigma}}{{N_{s_1}}}
\prod_{i=2}^N 
p_{s_i}(s_{i+1}|s_{i-1}).
\end{equation}
The index $\sigma$ denotes the direction needed to go from
$s_1$ to $s_{2}$,
$P_{s_1}$ is the probability for choosing site $s_1$ as the starting
point and $P_{w}^e$ is the probability for erasing the worm $w$ after
construction. $p_{s_i}(s_{i+1}|s_{i-1})$ is the conditional probability
for continuing to site $s_{i+1}$, at site $s_i$, given that the worm is
coming from $s_{i-1}$.
If the worm $w$ has been accepted we have to consider the
probability for reversing the move. That is, we consider the probability 
for constructing an anti-worm
$\bar w$ annihilating the worm $w$. We have:
\begin{equation}
P_{\bar w} = 
P_{\bar s_1}(1-P_{\bar w}^e)
\frac{{\bar A}_{s_1}^{\bar \sigma}}{{\bar N}_{\bar s_1}}
\prod_{i=N}^2
p_{s_i}(s_{i-1}|s_{i+1}).
\end{equation}
Here, the index $\sigma$ denotes the direction needed to go from
$s_1$ to $s_{N}$,
Note that, in this case the sites are visited in the opposite order,
$s_1,s_{N},\ldots,s_2$.
From Eq.~(\ref{eq:balance}) we have that
$p_{s_i}(s_{i+1}|s_{i-1})/p_{s_i}(s_{i-1}|s_{i+1})=
A^{\sigma_k}_{s_i}/{\bar A^{\sigma_l}_{s_i}}$. Since,
\begin{equation}
{A_{s_i}^{\sigma}}/
{\bar A_{s_{i}}^{\sigma}}=
\exp(-\Delta E_{s_i}^{\sigma}/K),\ i=1\ldots N,
\end{equation}
and since $P_{s_1}=P_{\bar s_1}$, we find:
\begin{equation}
\frac{P_w}
{P_{\bar w}}
=\frac{1-P_{w}^e}{1-P_{\bar w}^e}
\frac{\bar N_{\bar s_1}}{N_{s_1}}
\exp(-\Delta E_{\rm Tot}/K).
\end{equation}
where $\Delta E_{\rm Tot}$ is the total energy difference between a
configuration with and without the worm $w$ present. With our definition
of $P^e$ we see that
$(1-P^e(w))/(1-P^e(\bar w))=N_{s_1}/{\bar N_{s_1}}$ and
it follows that:
\begin{equation}
\frac{P_w}{P_{\bar w}}=
\exp(-\Delta E_{\rm Tot}/K).
\end{equation}
For a worm of length $N$ there are $N$ starting sites that will yield the
same final configuration. The above proof shows that for each of the starting sites
there exists an anti-worm, $\bar w$, such that $P_w=\exp(-\Delta E_{\rm Tot}/K)P_{\bar w}$.
Hence, if we by $\mu$ denote the configuration without the worm and $\nu$ the configuration
with the worm and furthermore let $P_w(s_i)$ denote the probability of building the worm $w$ starting
from site $s_i$, we see that:
\begin{eqnarray}
\frac{P(\mu\to\nu)}{P(\nu\to\mu)}& &
=\frac{\sum_i^N P_w(s_i)}{\sum_i^N P_{\bar w}(s_i)}\nonumber\\
& &=\frac{\sum_i^N P_{\bar w}(s_i)}{\sum_i^N P_{\bar w}(s_i)}\exp(-\Delta E_{\rm Tot}/K)\nonumber\\
& &=\exp(-\Delta E_{\rm Tot}/K).
\end{eqnarray}
Ergodicity is proved the same way as for the undirected algorithm as the worm can
perform local loops and wind around the
lattice in any direction, as in the conventional algorithm.

%%%%%%%%%%%%%%%%%%%%%%%%%%%%%%%%%%%%%%%%%%%%%%%%%%%%%%%%%%%%%%%%%%%%%%%%%%
%%%%%%%%%%%%%%%%%%%%%%%%%%%%%% Classical Algo %%%%%%%%%%%%%%%%%%%%%%%%%%%%
%%%%%%%%%%%%%%%%%%%%%%%%%%%%%%%%%%%%%%%%%%%%%%%%%%%%%%%%%%%%%%%%%%%%%%%%%%

\subsection{The Classical Worm algorithm}
\label{sec:classical}
Prokof'ev and Svistunov~\cite{Prokofev.Classical} have proposed a very elegant
way of performing Monte Carlo simulations on the high temperature expansion of
classical statistical mechanical models using worm algorithms.
In order to distinguish between the algorithms we call this algorithm the classical worm
algorithm. In a recent study~\cite{Prokofev.QR} these authors have performed simulations
on the quantum rotor model in the link-current representation, Eq.~(\ref{eq:hV}).
Due to the divergenceless constraint, the classical worm algorithm is in this case quite
close to the geometrical worm algorithm proposed previously in Ref.~\onlinecite{Alet03}
and not directly related to the high temperature expansion of this model.
Recasting their algorithm in the same framework used above we outline our understanding
of their algorithm below for comparison:
\begin{enumerate}
\item  Choose a random initial site $s_1=({\bf r}_1,\tau_1)$ in the space-time lattice.
\item For each of the directions $\sigma=\pm x,\pm y,\pm \tau$, calculate the probabilities
$A^\sigma_{s_i}$ with $A^\sigma_{s_i}=\min(1,\exp(-\Delta E_{s_i}^{\sigma}/K)),$ $
\Delta E_{s_i}^{\sigma}=E'^\sigma_{s_i}-E_{s_i}^{\sigma}$.\label{iter5}
\item With uniform probability choose a direction $\sigma$.
\label{loop5}
\item With probability $A^\sigma_{s_i}$ accept to go in the direction $\sigma$, and with
probability $1-A^\sigma_{s_i}$ go to \ref{loop5}.
\item Update $J^\sigma_{s_i}$ for the direction chosen and move the worm to the new lattice
site $s_{i+1}$.
\item If $s_i\ne s_1$ go to \ref{iter5}.
\item If $s_i=s_1$ go to 1 with probability $p_0$ and to \ref{loop5} with probability
$1-p_0$ ($p_0\in (0,1)$ and usually $p_0=1/2$). We use $p_0=1/2$ in the following.
\end{enumerate}
One advantage of this algorithm is its simplicity and the fact that a constructed
worm is always accepted, on the other hand this algorithm is not directed and steps
3-4 above are quite wasteful since in many cases the worm is not moved. This is
avoided in the geometrical worm algorithm at the price of occasionally having
to reject a complete worm. The geometrical worm algorithm, as described in the
preceding sections, should be straightforwardly applicable to the high temperature
expansion as it was done in Ref.~\cite{Prokofev.Classical} using the classical
worm algorithm. We expect that this would enhance the efficiency of the Monte Carlo sampling.

%%%%%%%%%%%%%%%%%%%%%%%%%%%%%%%%%%%%%%%%%%%%%%%%%%%%%%%%%%%%%%%%%%%%%%%%%%
%%%%%%%%%%%%%%%%%%%%%%%%%%%%%% Performances  %%%%%%%%%%%%%%%%%%%%%%%%%%%%%
%%%%%%%%%%%%%%%%%%%%%%%%%%%%%%%%%%%%%%%%%%%%%%%%%%%%%%%%%%%%%%%%%%%%%%%%%%

\section{Performance of the algorithms} \label{sec:auto}

Here we present results on autocorrelation times obtained with both directed
and undirected algorithms.  For the sake of brevity, we restrict ourselves
to the case $\mu=0$, where the critical point is known with high precision, and
where results on autocorrelations for the undirected worm algorithm have
already been published~\cite{Alet03}. All the results presented in this section
correspond to runs on cubic lattices of $10^7$-$10^8$ Monte Carlo worms for a
value of $K=0.333$, extremely close to the critical point (estimated as
$K_c=0.33305(5)$ in Ref.~\cite{Alet03}). In principle, simulations should be performed on
lattices of size $L\times L\times L_\tau$, but here since the dynamical exponent $z=1$ at $\mu=0$,
we can set $L_\tau=L$. We focus here on calculations of the
energy $E=\langle H \rangle$ and the stiffness $\rho$ defined as
\begin{equation}
\rho=\frac{1}{L^3}\langle (\sum_{{\bf r},\tau} J^x_{{\bf r},\tau})^2 \rangle
\end{equation}
where $L$ is the linear size of the lattice.

For the simulations with directed worms, we restrict ourselves to $|J| \leq
3$ for the tabulation of probabilities. Probabilities involving higher values of $|J|$
were calculated during the construction of the worm. Such configurations were found to
be exceedingly rare.

For the case at hand, only $1\%$ of the ''scattering'' matrices $P_{s_i}$
contained diagonal elements corresponding to a non-zero back-tracking
probability.  Moreover, these back-tracking (bounce) processes were found to
occur for very unlikely configurations. The acceptance rate, $1-P^e$, is very
high for both algorithms at $K_c$ (around $98\%$ for undirected worms and
$97\%$ for directed worms for all lattice sizes). For the classical worms, all
worms are accepted due to the nature of the algorithm.  However, we found that
many proposed attempts at changing one link were refused
(more than $60\%$ in our simulations).

\begin{figure}
\includegraphics[width=8cm]{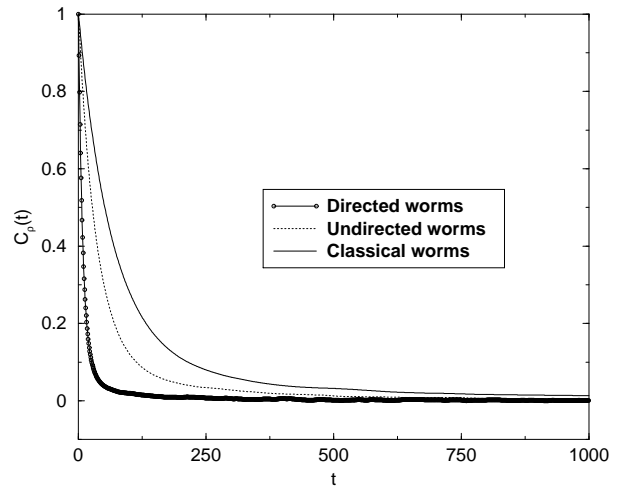}
\caption{Auto-correlation function of stiffness versus Monte Carlo time
  (defined by one worm construction - see text -) for
  $L=56$ at $K=0.333$ for directed (circle), undirected (dotted line) and
  classical (solid line) algorithms.}
\label{fig:auto.function}
\end{figure}

In figure~\ref{fig:auto.function}, we present the autocorrelation function of
stiffness for a lattice size $L=56$, for directed, undirected and classical worms. The
autocorrelation function $C_{\cal O}(t)$ of an observable ${\cal O}$ is defined
in the standard way:
\begin{equation}
 C_{\cal O}(t)=\frac{\langle {\cal
O}(t){\cal O}(0)\rangle - \langle {\cal O} \rangle^2}{\langle {\cal O}^2
{\rangle - \langle {\cal O} \rangle}^2}
\end{equation}
 where $\langle \ldots \rangle$ denotes statistical average and $t$ is the MC
time, measured in the number of constructed worms (accepted or not).  We
clearly see in figure~\ref{fig:auto.function} that the directed worm algorithm
is more efficient at decorrelating the data than undirected and classical
worms, the latter having the longest autocorrelation times.

Now we define the autocorrelation time $\tau_{\cal O}$ of an observable $\cal
O$. In Ref. \cite{Alet03}, $\tau_{\cal O}$ was defined as the greater time of a
double-exponential fit of the autocorrelation function.  Here we use a
much simpler definition, independent of any  fitting
procedure:  $\tau_{\cal O}$ is defined as the time where
the normalized autocorrelation function decrease below a
threshold $t_{\cal O}$. We can use different thresholds for different
observables $\cal O$. Since for small lattices and especially for directed
worms, autocorrelation times are small, and since $C_{\cal O}(t)$ is known only
for discrete values of $t$, $\tau_{\cal O}$ is determined by a simple linear interpolation
between the two times surrounding the threshold. It is important
to note that the values of the autocorrelation time depends on
the threshold $t_{\cal O}$, but the dependence on lattice size of these
autocorrelation times should not change as long as $t_{\cal O}$ is small
enough. Error bars on $t_{\cal O}$ have been estimated by slightly changing
the threshold, by an amount in between $2\%$ and $5\%$ in this work.

\begin{figure}
\includegraphics[width=8cm]{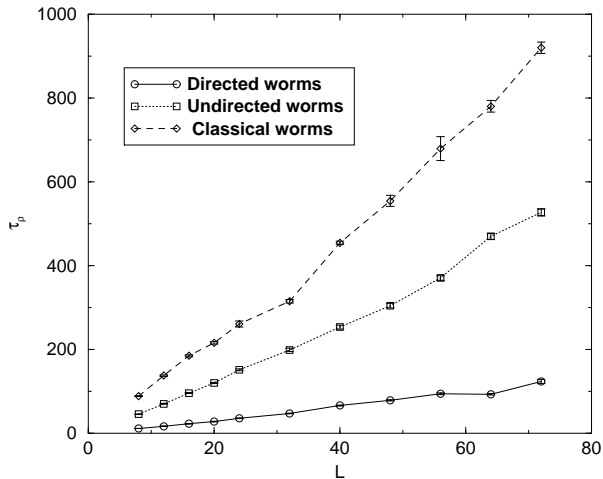}
\caption{Auto-correlation times of the stiffness, $\rho$, for directed, undirected and classical algorithms versus lattice size L.
Shown are the raw auto-correlations times, {\it before} rescaling to take into account the computational effort expended.}
\label{fig:auto.time.stiff}
\end{figure}

\begin{figure}
\includegraphics[width=8cm]{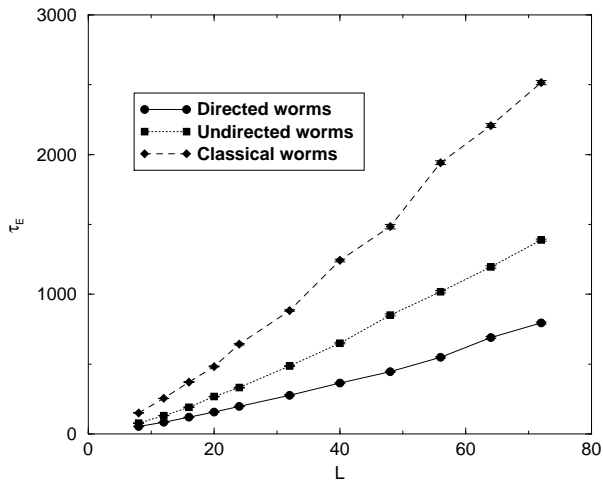}
\caption{Auto-correlation times of energy $E$ for the three algorithms versus lattice size L. Shown are the raw auto-correlations times,
{\it before} rescaling to take into account the corresponding computational effort expended.}
\label{fig:auto.time.en}
\end{figure}

Using the above mentioned determination of autocorrelation times, we extract
autocorrelation times of the stiffness $\rho $ and the energy $E$ for
directed, undirected and classical worms. The threshold was set the same for all algorithms
when comparing the same quantity: we used through this work $t_\rho=0.02$ for
the stiffness and $t_E=0.05$ for the energy. Scaling of these times with the
lattice size is shown in figure~\ref{fig:auto.time.stiff} for stiffness and in
figure~\ref{fig:auto.time.en} for energy. It can be seen that whereas
autocorrelation times grow approximatively linearly with lattice size for
all algorithms, the slope is significantly smaller for the directed worm algorithm.

It is clear from these results that the directed algorithm significantly
reduces the autocorrelation times. However, the average size of the directed worms could
be larger, and hence on average consume more computational
time. For all algorithms the computational effort is linearly proportional to
the length of the worm.  To make an honest comparison, we therefore have
to multiply the autocorrelation times by the number of attempted changes per
link, which we define as $\langle w \rangle /(3L^3)$, where $\langle w \rangle$
is the mean worm size (the mean number of links the worm has attempted to visit), $L$ the
lattice size preceded by an irrelevant factor indicating that there are $3$
links per site.  For the classical worms, the mean worm size $\langle w
\rangle$ is defined as the {\it total} number of proposed attempts (step 4 in
the pseudo-code presentation in section~\ref{sec:classical}).  In order to make
an un-biased comparison of the three algorithms it is here necessary to
include the updates refused during the construction of the classical worms in
the definition of $\langle w \rangle$.

\begin{figure}
\includegraphics[width=8cm]{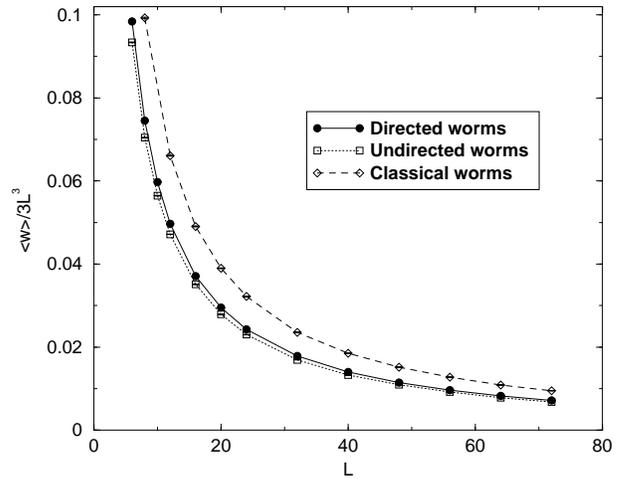}
\caption{Mean size $\langle w \rangle$ divided by $3L^3$ versus 
lattice size $L$ for directed, undirected and classical worms.}
\label{fig:Worm.Size}
\end{figure}

As mentioned, the computational effort (the CPU time) is linear in  $\langle w
\rangle$ for all algorithms.  An equivalent rescaling was used in
Ref.~\cite{Alet03} in order to make a fair comparison with the local algorithm.
In Fig.~\ref{fig:Worm.Size} is shown the mean worm size $\langle w \rangle$
(divided by $3L^3$) for the three algorithms versus lattice size, corresponding
to the average fraction of the total number of links occupied by the worm. In
both cases, we see that this fraction decreases with $L$. We also note that the
classical worms are longer than in the other proposed algorithms, which will
result in larger autocorrelation times. Directed and undirected worms are
almost of the same size, with very slightly larger directed worms. The
corresponding effect on the value of rescaled effort (presented in the next
paragraph) will be small when comparing autocorrelation times for those
algorithms, however we wish to keep it present for a more fair analysis.

\begin{figure}
\includegraphics[width=8cm]{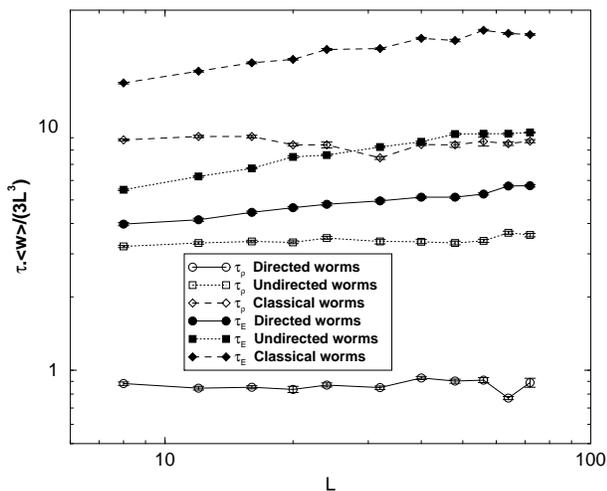}
\caption{Auto-correlation times of stiffness $\rho$ and energy $E$ for the three presented algorithms
versus lattice size L.
These auto-correlations times are {\it rescaled} auto-correlation times  where the computational effort is
taken into account.}
\label{fig:auto.time2}
\end{figure}

Having discussed the behavior of $\langle w \rangle $ we now take into account
the computational effort used to construct the worm by rescaling the
auto-correlation times by $\langle w \rangle /(3L^3)$.  We show in
figure~\ref{fig:auto.time2} the rescaled auto-correlation times for the three
algorithms. We find that the auto-correlation times per link stay reasonably
small for all algorithms, but the directed algorithm clearly gives better
results, with auto-correlation times smaller by a factor around $4$ ($1.5-1.7$)
for the stiffness (energy) with respect to the undirected algorithm, and a
factor around $10$ ($4$) with respect to the classical worm for the largest
sizes. The fact that both algorithms are more efficient at decorrelating the
stiffness than the energy seems to indicate that the worms couple more
effectively to ``winding modes'', from which the stiffness is uniquely
determined, than to simple local modes which determine the energy. With the
same argument, we can see that directed worms are more efficient at updating
winding modes than undirected or classical worms.

\begin{figure}
\includegraphics[width=8cm]{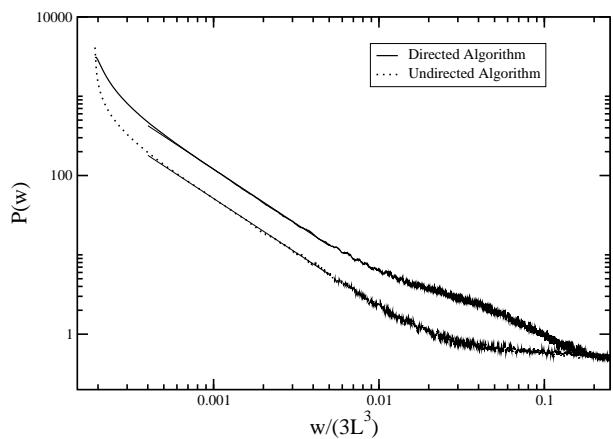}
\caption{The probability density, $P(w)$ for generating a worm occupying a fraction of $w/3L^3$
of the lattice, as a function of $w/3L^3$ for the directed and un-directed algorithms.
Shown are results for a lattice of linear size $L=56$ at $K=K_c$. The solid lines indicate power-law fits to the data.}
\label{fig:wsize}
\end{figure}

The actual distribution of the size of the worms generated, $P(w)$ is also of
interest.  In Fig.~\ref{fig:wsize} we show results for the probability density,
$P(w)$ for generating a worm occupying a fraction of $w/3L^3$ of the lattice,
as a function of $w/3L^3$ for the directed and un-directed algorithms.  The
classical worm algorithm has a distribution identical to the one shown for the
un-directed algorithm.  Clearly, the directed worms have a somewhat broader
distribution but for both algorithms the distribution follows a power-law form
$P(w)\sim w^{-\alpha}$ with $\alpha\sim 1.37$. The power-law behavior is to be
expected since the simulations were performed at the critical point. Away from
the critical point we have verified that the initial power-law form crosses
over to an exponential behavior at large arguments.

To summarize, we find that in all cases, rescaled auto-correlation times stay
almost constant with the lattice size, but could also be fitted with a very
small power-law or logarithm, showing an almost complete elimination of
critical slowing down. All in all, the main result of this section is that
directed worms produce less correlated data (smaller autocorrelation times),
even if the scaling is good for all the three (directed, undirected and
classical) algorithms. We also note that the geometrical worm algorithms 
perform better than the classical worm algorithm.

The fact that both directed and undirected algorithms have almost the
same scaling of the rescaled autocorrelation time with $L$,
seems also to be observed for the directed loop Quantum Monte Carlo
cluster algorithms~\cite{Sylju}.

%%%%%%%%%%%%%%%%%%%%%%%%%%%%%%%%%%%%%%%%%%%%%%%%%%%%%%%%%%%%%%%%%%%%%%%%%%
%%%%%%%%%%%%%%%%%%%%%%%%%%%%%% Green functions  %%%%%%%%%%%%%%%%%%%%%%%%%%
%%%%%%%%%%%%%%%%%%%%%%%%%%%%%%%%%%%%%%%%%%%%%%%%%%%%%%%%%%%%%%%%%%%%%%%%%%

\section{The Correlation Functions}
\label{sec:cfunc}
\subsection{Measurements of correlation functions with worm algorithms}

\begin{figure}
\includegraphics[width=8cm]{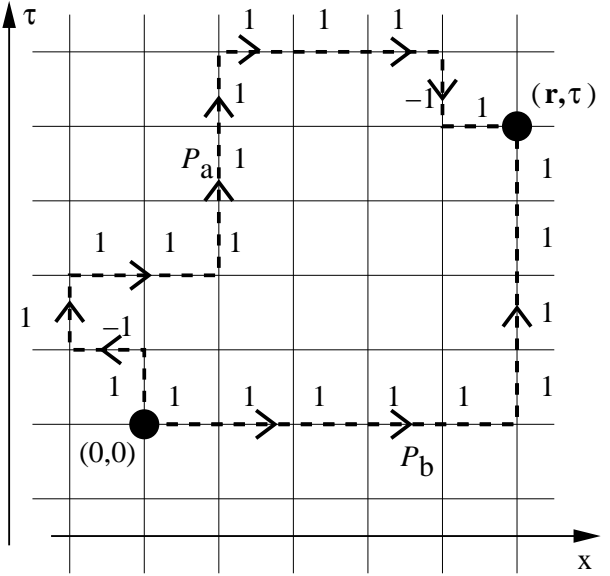}
\caption{Two possible paths, ${\cal P}_a$ and ${\cal P}_b$, for the
evaluation of $C({\bf r},\tau)$. When the path is going in the
$x,y,\tau$ direction a {\it particle} is propagated in the forward
direction corresponding
to an {\it increment} in the current. When the path
is going in $-x,-y,-\tau$ direction we propagate a {\it hole}
in the forward direction corresponding to a {\it decrement} in the current.
The solid circles correspond to sites where
a {\it single} particle is created or destroyed.
}
\label{fig:corr}
\end{figure}

For the quantum rotor model, the correlation functions of interest
have the following form~\cite{Sorensen}:
\begin{equation}
C({\bf r,r'},\tau,\tau') =  \langle e^{i(\hat\theta_{\bf r}(\tau)-
\hat\theta_{\bf r'}(\tau'))}\rangle \ ,
\label{eq:thetacor}
\end{equation}
where the $\hat\theta$'s are operators for the phase
of the bosons, and  $e^{i\hat\theta_{\bf r}(\tau)} =
e^{\tau H}e^{i\hat\theta_{\bf r}}e^{-\tau H}$.
Due to translational invariance,
$C({\bf r,r'},\tau,\tau')=C({\bf r-r'},\tau-\tau')$.
Physically this corresponds to creating a particle at $({\bf r},\tau)$
and destroying it at $({\bf r'},\tau')$.
When this correlation function is mapped onto the link-current
representation the creation and destruction of the particle is
interpreted as a particle current going from $({\bf r},\tau)$
to $({\bf r'},\tau')$. As is evident from the definition of the
correlation function in Eq.~(\ref{eq:thetacor}) the value of the
correlation function can not depend on the specific path taken
from $({\bf r},\tau)$ to $({\bf r'},\tau')$ as long as we 
take into account the fact that going in the $x,y,\tau$ increases
the local current, whereas going in the $-x,-y,-\tau$ direction 
decreases the local current.
In the link current representation this correlation function
can be written~\cite{Sorensen} in the following way:
\begin{eqnarray}
& &C({\bf r},\tau)=
\langle 
\prod_{({\bf r}_i,\tau_i)\; \rm \in path}\nonumber\\
& &\exp \left\{ -\frac{1}{K} 
\left({\rm sign}(\sigma_i)\left(J_{( {\bf
r}_i,\tau_i)}^\nu-\delta_{\sigma_i,\pm\tau}\tilde\mu_{\bf r_i}\right)+\frac{1}{2}\right) 
\right\}
\ 
\rangle,\nonumber\\
\label{eq:cr}
\end{eqnarray}
where ``path'' is any path on the space-lattice 
connecting two points a distance $({\bf r},\tau)$ apart
and $\sigma_i$ is the direction needed to
go from $({\bf r_i},\tau_i)$ to $({\bf r_{i+1}},\tau_{i+1})$,
$\sigma_i=\pm x,\pm y, \pm\tau$. When going in the direction
$\sigma_i=x,y,\tau$ we propagate a {\it particle} and the correlation
function corresponds to {\it incrementing} the corresponding
link-variable by one. When going in the direction $\sigma_i=-x,-y,-\tau$
we propagate a {\it hole} in the $x,y,z$ direction and the correlation
function corresponds to 
{\it decrementing} the corresponding link-variable by one. This is
indicated in the expression Eq.~(\ref{eq:cr}) by ${\rm sign}(\sigma_i)$.
Furthermore, we only get a contribution from $\mu_{\bf r_i}$ whenever we
go in the $\tau-$direction and we take this into account by
$\delta_{\sigma_i,\pm\tau}$. 
If we define $J^{-x}_{(x,y,\tau)}=-J^{x}_{(x-1,y,\tau)}$ with analogous
definitions for the other directions we see that by incrementing and
decrementing the link-current variables in the above
manor $\sum_\sigma J^\sigma_{({\bf r},\tau}=0$ at all the sites 
between $({\bf r_i},\tau_i)$ and $({\bf r_{i+1}},\tau_{i+1})$.
The current is divergenceless at all the intermediary sites.
The sites $({\bf r_i},\tau_i)$ and $({\bf r_{i+1}},\tau_{i+1})$ will
have non-zero divergence with $\sum_\sigma J^\sigma_{({\bf r},\tau}=1$
corresponding to a site where a particle is created (or a hole
destroyed). A site with $\sum_\sigma J^\sigma_{({\bf r},\tau}=-1$ is
a site where a hole is created (or a particle destroyed).
In Fig.~\ref{fig:corr} we show two possible
paths ${\cal P}_a$ and ${\cal P}_b$ for the evaluation of the correlation
function $C({\bf r},\tau)$.
As usual, $C({\bf r},\tau)=C({\bf r}+L,\tau+L_\tau)$ but
$C({\bf r},\tau)$ is in general not equal to $C({\bf r},-\tau)$.

Previous work~\cite{Sorensen,Kisker} have attempted to calculate the
correlation function by evaluating the thermal expectation value
in Eq.~(\ref{eq:cr}) along a straight path from $({\bf r},\tau)$
to $({\bf r'},\tau')$. Although formally correct, this method fails
for large arguments of the correlation function due to the fact that
for a given configuration of the link-variables roughly only {\it one} specific
path between $({\bf r},\tau)$ and $({\bf r'},\tau')$ will yield a
contribution of order 1.

The geometrical worm algorithm allows for a much more efficient way
of evaluating the correlation functions. In essence, before the worm
returns to the starting site, the path of the worm corresponds precisely
to the creation of a particle at site $s_1$ and the destruction at the
current site $s_i$ with a current going between the two sites. 
This is precisely the Greens function that we want
to calculate. More precisely we extend Eq.~(\ref{eq:cr}) to include a
summation over all possible paths:
\begin{eqnarray}
& &C({\bf r},\tau)=\frac{1}{N_{\cal P}}\sum_{\cal P}
\langle 
\prod_{({\bf r}_i,\tau_i)\; \rm \in {\cal P}}\nonumber\\
& &\exp \left\{ -\frac{1}{K}
\left({\rm sign}(\sigma_i)\left(J_{( {\bf
r}_i,\tau_i)}^\nu-\delta_{\sigma_i,\pm\tau}\tilde\mu_{\bf r_i}\right)+\frac{1}{2}\right) 
\right\}
\
\rangle.\nonumber\\
\label{eq:crall}
\end{eqnarray}
Here ${\cal P}$ is a path for the correlation function and $N_{\cal P}$
is the number of paths included in the sum. Since the geometrical worm algorithm
generates paths between $({\bf r},\tau)$ and $({\bf r}_n,\tau_n)$ with the
correct exponential factor (except for a multiplicative constant) it is now
easy to calculate the correlation functions.

Suppose that we, by using either the directed or undirected worm
algorithm, have reached the equilibrium configuration $\mu$. The
probability for, during the construction of a worm starting  at site 
$s_1=({\bf r}_1,\tau_1)$, creating a current $j$ that reaches 
$s_n=({\bf r}_n,\tau_n)\neq s_1$ is given by:
\begin{equation}
P(j;\mu\to\mu')=P(s_1)\prod_{i=1}^{n-1}\frac{A_{s_i}^\sigma}{N_{s_i}}
\end{equation}
for the undirected algorithm. For the directed algorithm we have:
\begin{equation}
P(j;\mu\to\mu')=P(s_1)\frac{A_{s_1}^\sigma}{N_{s_1}}
\prod_{i=2}^{n-1}p_{s_i}(s_{i+1}|s_{i-1}).
\end{equation}
If we call the resulting state $\mu'$ we can calculate the probability
for, starting from $\mu'$, creating an anti-current, $\bar j$, going from $s_n$
to $s_1$. We find for the undirected algorithm:
\begin{equation}
P(\bar j;\mu'\to\mu)=P(s_n)\prod_{i=n}^2\frac{\bar
A_{s_i}^\sigma}{N_{s_i}},
\end{equation}
and for the directed algorithm:
\begin{equation}
P(\bar j;\mu' \to\mu)=P(s_N)\frac{\bar A_{s_n}^\sigma}{\bar N_{s_n}}
\prod_{i=n-1}^{2}p_{s_i}(s_{i-1}|s_{i+1}).
\end{equation}
In both cases we see that
\begin{eqnarray}
& &\frac{P(j;\mu\to\mu')}{P(\bar j;\mu'\to\mu)}=
\frac{\bar N_{s_n}}{N_{s_1}}
\prod_{({\bf r}_i,\tau_i)\; \rm \in {\cal P}}\nonumber\\
& &\exp \left\{ -\frac{1}{K}
\left({\rm sign}(\sigma_i)\left(J_{( {\bf
r}_i,\tau_i)}^\nu-\delta_{\sigma_i,\pm\tau}\tilde\mu_{\bf r_i}\right)+
\frac{1}{2}\right) \right\}.\nonumber\\
\end{eqnarray}
Hence, we see that for both algorithms the intermediate states generated during the construction
of the worm follows precisely the distribution needed apart from the factor
${\bar N}_{s_n}/{N_{s_1}}$.
It follows that whenever a worm reaches a point a distance
$({\bf r},\tau)$ away from the initial point it contributes 
a factor of $N_{s_1}/{\bar N_{s_n}}$ to the correlation function
of argument $({\bf r},\tau)$. Note that it follows from the above proof
that {\it all} worms, even the ones that are finally rejected, have to
be included in the calculation of the Greens functions.
Per definition $C(0,0,0)\equiv C(L,L,L_\tau)\equiv 1$.

\subsection{Results}

The above procedure is straight forward to implement. Suppose we want to calculate the Greens functions
for a $d+1$ dimensional system with $d=2$. Since the two space directions are equivalent by symmetry
it is only necessary to calculate $C(x,\tau)$. This is easily done by keeping track of the position
of the worm during construction. If the relative position of the worm with respect to its starting point $s_1$,
is denoted by $(x_r,y_r,\tau_r)$, when the worm has reached site $s_n$, we add $N_{s_1}/{\bar N_{s_n}}$ 
to $C(x_r,y_r,\tau_r)$. This can be done with very
little computational effort and since an enormous amount of worms are generated during the simulation
extremely good statistics can be obtained for $C(x_r,y_r,\tau_r)$ by averaging
over the worms (which cannot be achieved with the local algorithm). As mentioned, in order not to bias the
calculation, even worms that are eventually rejected should be included
for a correct calculation of the Greens functions.
\begin{figure}
\includegraphics[width=8cm]{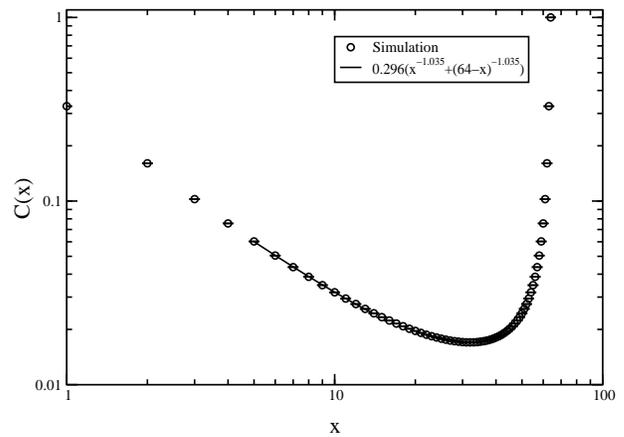}
\caption{The Greens function $C(x)$ for a system of size $L^3, L=64$ at $K=K_c=0.33305$, $\mu=0$,
as a function of $x$. The solid line indicates a power-law fit of the form $0.296(x^{-1.035}+(64-x)^{-1.035}).$}
\label{fig:cx}
\end{figure}
In Figure~\ref{fig:cx} we show results for the Greens function as a function of $x$ for a system
of size $L^3, L=64$. For this simulation the directed algorithm was used with a total number of worms
equal to $1.5\times 10^8$.
It is easy to obtain extremely small error bars on the Greens functions even for
very large system sizes. For the results shown in Fig~\ref{fig:cx} $\mu=0$ and by symmetry $C(\tau)$ is
identical to $C(x)$. From scaling relations~\cite{Fisher89c} $C(r)$ is expected to decay as $r^{-(d-2+z+\eta)}$ where $z$ is the dynamical critical
exponent. 
With $\mu=0$, $z=1$ we find $C(r)\sim r^{-(1+\eta)}$. Fitting
to this form we find $\eta=0.035(5)$.
The obtained critical exponents are in excellent agreement with previous work~\cite{Sorensen}
and more recent high-precision estimates for the critical exponents of the 3d $XY$ model~\cite{Dukovski}.

It would be of much interest to calculate $C({\bf r},\tau)$ for $\mu\neq 0$ using
this method. Such calculations are currently in progress~\cite{muquarter}.

%%%%%%%%%%%%%%%%%%%%%%%%%%%%%%%%%%%%%%%%%%%%%%%%%%%%%%%%%%%%%%%%%%%%%%%%%%
%%%%%%%%%%%%%%%%%%%%%%%%%%%%%% Conclusion %%%%%%%%%%%%%%%%%%%%%%%%%%%%%%%%
%%%%%%%%%%%%%%%%%%%%%%%%%%%%%%%%%%%%%%%%%%%%%%%%%%%%%%%%%%%%%%%%%%%%%%%%%%

\section{Summary and discussion}

We have proposed a directed worm algorithm for the quantum rotor
model. This algorithm is an improvement of the ''undirected'' algorithm
presented in~\cite{Alet03}. It has been shown that by adjusting the degrees
of freedom left in the detailed balance condition, one can construct a more
efficient algorithm by minimizing the back-tracking (bounce) probability for the worm to erase itself.
The minimal probabilities can be found by solving a linear programming problem
subject to a few well-defined constraints.
A proof of detailed balance for the directed case has also been presented.
The directed and un-directed algorithms are identical except for the fact that
appropriately defined local probabilities $p_\sigma$ for moving the worm through
the lattice are chosen in an optimal manner for the directed algorithm. Hence,
only a very limited amount of additional programming has to be done to implement the
directed algorithm.

These central ideas for this directed algorithm can be straightforwardly applied to directed QMC loop
algorithms~\cite{Sylju} and one can avoid an analytical calculation for each new model
where one wants to implement a directed algorithm. More generally speaking, we
believe that the framework presented here could be useful for constructing new
algorithms for other models, for example classical spin
models~\cite{Hitchcock}.

We have shown the superiority of the directed algorithm as compared to the
undirected one and to the approach ("classical worms") proposed in~\cite{Prokofev.Classical}
by calculating autocorrelation times of different observables
near a critical point. Whereas the computational gain is not as drastic as
when passing from a local update algorithm to a worm
algorithm~\cite{Alet03,AletPhD}, we showed that one gains a factor ranging from
$1.5$ to $10$ (depending on the quantity and on the comparison) for the simulations considered here.
We did not try to estimate autocorrelation exponent $z$ for the algorithms,
because in all cases, it is small (as can be seen in
figure~\ref{fig:auto.time2}) and it would be hard to determine with high
precision. Looking at the data, it is likely that values of $z$ for all
algorithms are the same or quite close. A logarithmic dependence of $\tau$ on
$L$, indicating $z=0$, cannot also be excluded.

In this paper, we have also derived an efficient way of measuring correlation
functions during the worm constructions. This feature is similar to other worm
algorithms~\cite{Sandvik,Prokofev}, but here we show, including analytical arguments,
that it also works for directed worms. The situation for directed QMC loop algorithms~\cite{Sylju} is
less certain, even if some results were recently presented in~Ref.\cite{roscilde}.

The directed worm algorithm could be specially useful to
study the transition for a non-commensurate value of the chemical potential in
the pure quantum rotor model or for the disordered case, where very strong finite size effects have been
identified~\cite{AletPhD,muquarter,Prokofev.QR}.

\begin{acknowledgments}
We thank M.~Troyer  for useful discussions and J.~Asikainen for a careful
reading of the manuscript. This work is
supported by the NSERC of Canada, the SHARCNET computational initiative and by
the Swiss National Science Foundation.
\end{acknowledgments}

% If in two-column mode, this environment will change to single-column
% format so that long equations can be displayed. Use
%\begin{widetext}
%\end{widetext}

% figures should be put into the text as floats.
% Use the graphics or graphicx packages (distributed with LaTeX2e)
% and the \includegraphics macro defined in those packages.
% See the LaTeX Graphics Companion by Michel Goosens, Sebastian Rahtz,
% entire page. There is no need to do explicit centering.

% \begin{figure}
% \end{figure}

% Surround figure environment with turnpage environment for landscape
% figure
% \begin{figure}
% \end{figure}

\end{document}